\documentclass[12pt]{article}
\usepackage{hyperref}
\usepackage{amssymb,amsmath,graphicx,latexsym}

\begin{document}
\title{A note on the effect of the cosmological constant on the bending of light}
\author{Fay\c{c}al Hammad\thanks{fayhammad@gmail.com}\\
\emph{\small Universit\'{e} A.Mira. Route Targa Ouzemmour, Bejaia 06000. Algeria,}}
\date{}
\maketitle
\begin{abstract}
We take another look at the equations behind the description of light bending in a Universe with a cosmological constant. We show that even within the impact parameter entering into the photon's differential equation, and which is defined here with exclusive reference to the beam of light as it bends around the central mass, lies the contribution of the cosmological constant. The latter is shown to inter in a novel way into the equation. When the latter is solved our approach implies, beyond the first two orders in the mass-term and the lowest-order in the cosmological constant, a slightly different expression for the bending angle from what is previously found in the literature.
\end{abstract}

\begin{quote}

\end{quote}



\section{Introduction}\label{sec:1}
It is by now firmly established that we live in an accelerated expanding Universe \cite{1,2,3}. It is then of primordial importance to reconsider every known dynamics taking place in the Universe in the light of this fact. The natural place to start is our own solar system, a system that served well in testing general relativity at great precisions \cite{4}.

Since the actually observed Universe is modeled quite well by introducing a cosmological constant $\Lambda$ (see e.g. \cite{5}), any investigation of the consequences of the cosmic expansion amounts to find any contribution of $\Lambda$ in shaping the studied phenomenon.

It is by now relatively well agreed upon the fact that a cosmological constant does effect the bending of light around a central mass like the Sun. In the past, however, researchers have obtained the counterintuitive result that this is not the case by deriving the photon's differential equation as it travels around the mass all the way to the observer \cite{6,7,8}. The first-order differential equation obtained there does not indeed contain any explicit contribution of the cosmological constant $\Lambda$. What recent authors in \cite{9} have shown is that the obtained equation whose solution describes the photon's trajectory should not be considered as the definitive answer for one cannot rely solely on the trajectory that merely gives the shape of the path because one still needs an input from geometric considerations in order to complement the analysis. Namely, in order to be able to extract at the end the measured angle of deflection one must in addition use the information contained in the curved space whose shape is also related to the cosmological constant. (See also \cite{10,11,12,13,14}.)

In the present note we take another look at the same differential equation and argue that within this equation lies actually the effects of the cosmological constant, albeit hidden within the impact parameter. Indeed, although in \cite{9} it was shown at what stage the effects of the cosmological constant inter into the problem, one still cannot but wonder how it is that the photon's trajectory remains insensitive to the fact that space-time is expanding. A work in this direction can be found in \cite{11,14}. There, however, the authors took as in \cite{9} a different definition for the impact parameter from that of \cite{8}. In \cite{9} the impact parameter is defined with reference to the shortest distance of approach of the straight undeflected beam of light from the central mass when putting the latter to zero, whereas the impact parameter in \cite{8} is deduced exclusively from the deflected beam as it bends in the presence of the mass. Adopting the former approach does not explain prior to solving the differential equation why the latter appears independent of the cosmological constant when the impact parameter solely refers to the deflected beam. In the present note we shall see that exclusive reference to the deflected beam permits not only to display in a novel way the effects of $\Lambda$ at the differential equation level but to obtain at the end a slightly different approximation for the bending angle as well.

\section{The differential equation, the cosmological constant, and the bending angle}\label{sec:2}

The metric around a mass $M$ in an expanding Universe due to a cosmological constant $\Lambda$ is described by the Schwarzschild-de Sitter-Kottler metric \cite{15} that reads, using units in which $G=c=1$ \cite{16},
\begin{equation}\label{1}
\mathrm{d}s^2=-\left(1-\frac{2M}{r}-\frac{\Lambda}{3}r^{2}\right)\mathrm{d}t^{2}+\left(1-\frac{2M}{r}-\frac{\Lambda}{3}r^{2}\right)^{-1}\mathrm{d}r^{2}+r^{2}\mathrm{d}\Omega^{2},
\end{equation}
where $\mathrm{d}\Omega^{2}=\mathrm{d}\theta^{2}+\sin^{2}\theta\mathrm{d}\varphi^{2}$. From this metric one extracts the Lagrangian $\mathcal{L}$ for a point particle moving in the equatorial plane $\theta=\pi/2$ \cite{16}:
\begin{equation}\label{2}
\mathcal{L}=-\alpha\dot{t}^{2}+\alpha^{-1}\dot{r}^{2}+r^{2}\dot{\varphi}^{2},
\end{equation}
where a dot over a letter denotes the differentiation $\mathrm{d}/\mathrm{d}s$. Also, in order to avoid cumbersome formulas below, we introduced as is customary the following function of $r$
\begin{equation}\label{3}
\alpha(r)=\left(1-\frac{2M}{r}-\frac{\Lambda}{3}r^{2}\right).
\end{equation}
Using the Euler-Lagrange equations
\begin{equation}\label{4}
\frac{\mathrm{d}}{\mathrm{d}s}\left(\frac{\partial\mathcal{L}}{\partial \dot{x}^{\mu}}\right)-\frac{\partial\mathcal{L}}{\partial x^{\mu}}=0,
\end{equation}
one extracts the geodesic equations. The Lagrangian being independent of the variables $t$ and $\phi$ we have at once the following two constants of motion, representing the total energy and angular momentum respectively \cite{16},
\begin{equation}\label{5}
E=\alpha\dot{t}\quad\mathrm{and}\quad L={r}^2\dot{\varphi}.
\end{equation}
Putting $\mathcal{L}=0$ in (\ref{2}) for a beam of light and then using the identities (\ref{5}) we also find \cite{8}
\begin{equation}\label{6}
\left(\frac{\mathrm{d}r}{\mathrm{d}\varphi}\right)^{2}=r^{4}\left(\frac{1}{b^{2}}-\frac{\alpha}{r^{2}}\right),
\end{equation}
where we introduced as it is also customary the impact parameter $b=L/E$. This parameter is actually related to the nearest coordinate distance $R_{0}$ of approach of the beam of light to the masse $M$ through the condition \cite{8}, $(\mathrm{d}r/\mathrm{d}\varphi)_{r=R_{0}}=0$. From (\ref{6}) one then deduces that $1/b^{2}=\alpha_{0}/R_{0}^{2}$, where $\alpha_{0}$ denotes the value of $\alpha(r)$ at the radius $r=R_{0}$. Substituting this back into (\ref{6}) the latter in fact transforms into a differential equation that seems to indicate that the cosmological constant does not intervene since it disappears completely from the final equation \cite{8}:
\begin{equation}\label{7}
\left(\frac{\mathrm{d}r}{\mathrm{d}\varphi}\right)^{2}=\left(\frac{\alpha_{0}}{R_{0}^{2}}-\frac{\alpha}{r^{2}}\right)r^{4}=\left(\frac{1}{R_{0}^{2}}-\frac{2M}{R_{0}^{3}}\right)r^{4}-r^{2}+2Mr.
\end{equation}

\begin{figure}[h]
\center\includegraphics[angle=-90, scale=0.5]{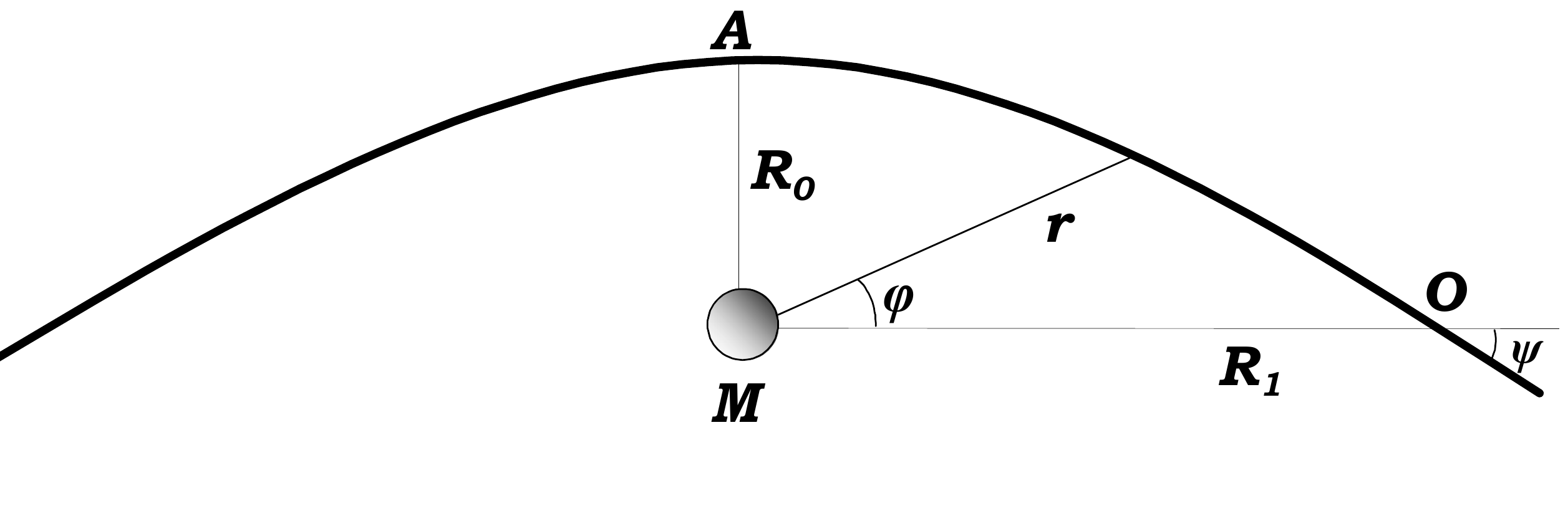}\\
  \caption{Light trajectory around a mass $M$. The point $A$ corresponds to the beam's nearest approach to the mass while the point $O$ corresponds to the position of the observer.}\label{Fig}
\end{figure}

Actually, the cosmological constant has also its word in defining the nearest distance of approach $R_{0}$ to the mass in order for the detected photons to achieve the detector under the observation angle. This can be seen as follows. First, take the ratio of the two constants of motion in (\ref{5}) at two distinct positions along the orbit of the light. Taking the first position to be the closest approach which is at $r=R_{0}$ while the second position to be at the location of the observer's laboratory at the coordinate distance $r=R_{1}$ from the central mass (see Fig.\ref{Fig}), and denoting by $\alpha_{1}$ the value of $\alpha(r)$ at $r=R_{1}$, we get
\begin{equation}\label{8}
\frac{1}{b^{2}}=\frac{\alpha_{0}}{R_{0}^{2}}=\frac{\alpha_{1}^{2}}{R_{1}^{4}\left[\left(\frac{\mathrm{d}\varphi}{\mathrm{d}t}\right)^{2}\right]_{r=R_{1}}}.
\end{equation}

Next, given that for a light beam we have $\mathrm{d}s=0$ we can extract from the metric (\ref{1}) the following identity at the equatorial plane $\theta=\pi/2$,
\begin{equation}\label{9}
\frac{\alpha^{2}}{r^{4}\left(\frac{\mathrm{d}\varphi}{\mathrm{d}t}\right)^{2}}=\frac{\alpha}{r^{2}}+\frac{1}{r^{4}}\left(\frac{\mathrm{d}r}{\mathrm{d}\varphi}\right)^{2}.
\end{equation}
Since this is an identity that is supposed to be valid everywhere along the photon's trajectory, it should also be satisfied at the laboratory's position at the radial distance $r=R_{1}$. On the other hand, as indicated in detail in \cite{9,10} the spatial metric at the equatorial plane $\theta=\pi/2$ being $\mathrm{d}\textbf{\textit{l}}^{2}=\alpha^{-1}\mathrm{d}r^{2}+r^{2}\mathrm{d}\varphi^{2}$, the tangent of the angle between the radial direction and the tangent to the photon's trajectory is given by $r\sqrt{\alpha}(\mathrm{d}r/\mathrm{d}\varphi)^{-1}$. Thus, at the very location where the measurements are taking place we have (see Fig. \ref{Fig})
\begin{equation}\label{10}
\left[\left(\frac{\mathrm{d}r}{\mathrm{d}\varphi}\right)^{2}\right]_{r=R_{1}}=\frac{R_{1}^{2}\alpha_{1}}{\tan^{2}\psi}.
\end{equation}
The bending angle is defined to be twice the angle $\psi$ \cite{9}.

Writing equation (\ref{9}) at the position $r=R_{1}$ and then substituting the above identity gives
\begin{equation}\label{11}
\frac{\alpha_{1}^{2}}{R_{1}^{4}\left[\left(\frac{\mathrm{d}\varphi}{\mathrm{d}t}\right)^{2}\right]_{r=R_{1}}}=\frac{\alpha_{1}}{R_{1}^{2}\sin^{2}\psi}.
\end{equation}
Finally, substituting this in (\ref{8}) we learn that
\begin{equation}\label{12}
\frac{1}{b^{2}}=\frac{\alpha_{0}}{R_{0}^{2}}=\frac{1}{\sin^{2}\psi}\left(\frac{1}{R_{1}^{2}}-\frac{2M}{R_{1}^{3}}-\frac{\Lambda}{3}\right).
\end{equation}

Thus, we see that at the end the shortest distance of approach $R_{0}$ of the observed photon to the mass depends not only on the cosmological constant, but also on the radial distance $R_{1}$ of the observer from the central mass as well as the angle of observation $\psi$. This might have been expected intuitively as well since one would be able to detect through the angle $\psi$ only those photons that have traveled through the path depicted in Fig. \ref{Fig}. In other words, the detected photons at the radial distance $R_{1}$ must have followed that geodesic which passed through the point $A$ at the specific distance $R_{0}$ from the mass in order for them to achieve the detector under the angle $\psi$.

The substitution of (\ref{12}) into (\ref{7}) permits to see explicitly that the cosmological constant does not really cancel away completely. Indeed, here is how the contributions of $\Lambda$ in (\ref{7}) get separated from those of the mass $M$:
\begin{equation}\label{13}
\left(\frac{\mathrm{d}r}{\mathrm{d}\varphi}\right)^{2}=\left(\frac{1}{R_{1}^{2}}-\frac{2M}{R_{1}^{3}}-\frac{\Lambda}{3}\cos^{2}\psi\right)\frac{r^{4}}{\sin^{2}\psi}-r^{2}+2Mr.
\end{equation}
\section{Finding the bending angle}\label{sec:3}
Now that we have obtained explicitly how and where the cosmological constant $\Lambda$ intervenes in the differential equation giving the photon's trajectory, all that remains to do to find the bending angle is to solve the equation for $r$ and then extract $\sin\psi$ from the solution. To solve (\ref{13}) it is customary to introduce the variable $u=1/r$ for which the equation takes the simpler form
\begin{equation}\label{14}
\left(\frac{\mathrm{d}u}{\mathrm{d}\varphi}\right)^{2}=\frac{1}{B^{2}}-u^{2}+2Mu^{3},
\end{equation}
where we introduced the constant
\begin{equation}\label{15}
\frac{1}{B^{2}}=\frac{1}{\sin^{2}\psi}\left(\frac{1}{R_{1}^{2}}-\frac{2M}{R_{1}^{3}}-\frac{\Lambda}{3}\right)+\frac{\Lambda}{3}.
\end{equation}
An equation identical to (\ref{14}) has already been introduced in \cite{14}. There, however, the constant $1/B^{2}$ is given by $1/B^{2}=1/b^{2}+\Lambda/3$ where $b$ is the impact parameter defined with respect to the undeflected beam of light. The solution to (\ref{14}) has also been given in \cite{14} up to the second-order in the mass-term; it reads,
\begin{equation}\label{16}
\frac{1}{r}=\frac{1}{B}\sin\varphi+\frac{M}{2B^{2}}(3+\cos2\varphi)+\frac{M^{2}}{16B^{3}}(37\sin\varphi+30(\pi-2\varphi)\cos\varphi-3\sin3\varphi).
\end{equation}

In order to extract the bending angle from this solution we simply evaluate the latter at the radius $r=R_{1}$, for which we also have $\varphi=0$, and then solve the resulting algebraic equation for $\sin\psi$. Thus, we first find $1/R_{1}=2M/B^{2}+15\pi M^{2}/8B^{3}+\mathcal{O}(M^{3})$ which yields, after substituting the complete expression (\ref{15}) for $1/B^{2}$ and multiplying both sides of the equation by $\sin^{2}\psi$,
\begin{equation}\label{17}
\sin^{2}\psi=P+\frac{Q}{\sin\psi}+\mathcal{O}(M^{3},\Lambda M^{2}),
\end{equation}
where we denoted by $P$ and $Q$ the following small quantities
\begin{eqnarray}\label{18}
P=\frac{2M}{R_{1}}\left(1-\frac{2M}{R_{1}}-\frac{\Lambda}{3}R_{1}^{2}\right),\nonumber
\\Q=\frac{15\pi M^{2}}{8R_{1}^{2}}\left(1-\frac{\Lambda}{2}R_{1}^{2}\right).
\end{eqnarray}
To extract $\sin\psi$ up to $\mathcal{O}(M^{2},\Lambda M)$ it is easier, given the form of the leading term in the quantity $P$, to just insert into (\ref{17}) the following expansion for $\sin\psi$,
\begin{equation}\label{19}
\sin\psi=c_{1}\sqrt{\frac{M}{R_{1}}}+c_{2}\frac{M}{R_{1}}+c_{3}\left(\frac{M}{R_{1}}\right)^{3/2}+c_{4}\Lambda\sqrt{MR_{1}^{3}}+\mathcal{O}(M^{2},\Lambda M).
\end{equation}
One then determines the coefficients $c_{i}$ by identifying at each order of the series the corresponding terms on the two sides of the equation. The final result for $\sin\psi$ is the following expression
\begin{eqnarray}\label{20}
\sin\psi=\sqrt{\frac{2M}{R_{1}}}+\frac{15\pi M}{32R_{1}}-\left(\sqrt{2}+\frac{675\pi^{2}}{2048\sqrt{2}}\right)\left(\frac{M}{R_{1}}\right)^{3/2}-\frac{1}{3\sqrt{2}}\Lambda\sqrt{MR_{1}^{3}}\nonumber
\\+\mathcal{O}(M^{2},\Lambda M).\qquad\qquad\qquad\qquad\qquad\qquad\qquad\qquad\qquad\qquad\qquad
\end{eqnarray}

At this point, we have actually recovered exactly the first- and second-order mass-terms as well as the first-order $\Lambda$-term that have already appeared in \cite{9}. Furthermore, as it can easily be checked, even the $\mathcal{O}(M^{3/2})$ mass-term appears when one uses the approach adopted in \cite{9} after reexpressing everywhere the impact parameter as it is defined there in terms of the radial distance $R_{1}$. What difference with respect to \cite{9} then, one might ask, does the present analysis bring apart from showing that $\Lambda$ might enter differently into the equations governing the bending of the photon's trajectory around the mass? The answer lies in the fact that here we have not derived an expression for $\tan\psi$ as is the case in \cite{9} but we obtained an expression for $\sin\psi$. To find the bending angle $2\psi$ one simply writes $\psi=\sin^{-1}(\sin\psi)$ and uses the above expression. As such, one will find discrepancies between our approach and that of \cite{9,10,11,12,13,14} when calculating the value of $\psi$ at $\mathcal{O}(M^{3/2})$ and beyond. This comes about due to the fact that $\sin^{-1}\epsilon=\epsilon+\epsilon^{3}/6+...$, whereas $\tan^{-1}\epsilon=\epsilon-\epsilon^{3}/3+...$.

As for the appearance of $\sin\psi$ in the final result instead of $\tan\psi$, it can be traced back to the appearance of $\sin\psi$ even within the impact parameter which in turn determines the shape of the photon's trajectory via the first integral (\ref{13}) of the geodesic equation.

\section{Summary}
We analyzed the geodesic followed by a beam of light as it bends around a massive object embedded in an expanding space-time. Trying to reveal the effect of a cosmological constant $\Lambda$ on the trajectory of the photons, and hence on their bending angle as they arrive at the observer, we found that $\Lambda$ does indeed contribute and this even at the first-order differential equation level. We found that it does so by bringing together the radial distance $R_{1}$ of the detector from the central mass, the value of the central mass $M$ itself, and the angle $\psi$ under which the photons are detected.

Taking into account these contributions in the differential equation and solving the latter gave us an algebraic equation in $\sin\psi$. When we extracted the latter we found that the terms previously obtained in the literature for $\tan\psi$ do in fact arise for $\sin\psi$. Hence, the resulting bending angle $2\psi$ agrees with the previous literature only up to $\mathcal{O}(M)$ and $\mathcal{O}(\Lambda\sqrt{M})$.

\end{document}